\documentclass[12pt]{article}
\usepackage{graphicx}
\usepackage{bm}
\usepackage{amssymb}
\begin{document}
\begin{titlepage}

\vspace{1cm}

\begin{center}{\Large {\bf Advantages of axially aligned crystals used
\\in positron production at future linear colliders }}\\
\vspace{1.0cm}
{\Large X.Artru$^a$, R.Chehab$^b$, M.Chevallier$^a$, and
V.Strakhovenko$^c$}
\footnote {Corresponding author, e-mail: v.m.strakhovenko@inp.nsk.su }\\
\vspace{1cm} {\it $^a$ IPN-Lyon, IN2P3/CNRS et Univ. Claude Bernard,
69622 Villeurbanne,
France\\ \vspace{0.5cm}
$^b$ LAL, IN2P3/CNRS et Univ. de Paris-Sud, BP 34-91898, Orsay cedex,
France\\
\vspace{0.5cm}$^c$ Budker-INP, 11 Ac.Lavrentyeva, 630090, Novosibirsk,
Russia }
\end{center}
\vspace{4.0cm}

\begin{abstract}

\noindent The characteristics of the electron-photon showers initiated by 2 to 10~GeV
electrons aligned along the $<111>$~-~axis of tungsten crystals are compared with those
for the amorphous tungsten . In this energy range, as known, the positron yield at the
optimal target thicknesses is larger in a crystal case only by several percent. However,
the amount of the energy deposition in a crystal turns out to be considerably (by 20 -
50$\% $) lower than in an amorphous target providing the same positron yield, while the
peak energy-deposition density is approximately of the same magnitude in the both cases.

\end{abstract}
\vspace{2cm}
\vspace{1cm}

\end{titlepage}
\newpage
\section{Introduction}

The formation of electromagnetic showers in aligned single crystals was actively studied
during the last decade. The first experimental investigation of such showers has been
performed in \cite {Rmed} at very high energy of incident electrons. Corresponding
theoretical studies were started with \cite{BKS87} where an analytic solution of the
problem was obtained, assuming that energies of all charged particles and photons
involved are very high. This limitation was surmounted in \cite{BKS95} by suggesting a
specific form of the radiation spectrum at axial alignment and performing corresponding
simulations. Using this approach, the results of \cite {Rmed} for Ge crystal had been
reproduced in\cite{BKS96}. The results of \cite{BKS95} are consistent with those of
\cite{Artru94} where another approach was used to obtain the radiation spectrum. In
\cite{BKS95},\cite{Artru94}, and \cite{BS99}, the shower characteristics, such as
spectral-angular distributions of photons and positrons as well as the amount of energy
deposition have been obtained depending on the kind and thickness of crystal targets.
Investigations of the specific shower formation give good grounds for the idea proposed
in \cite{Cheh89}, to substitute in a positron source an axially aligned crystal target
for an amorphous one, as the enhancement of the radiation intensity is maximal just at
the axial alignment. In further experiments (see [8-13]) using 1.2-10 GeV electrons
aligned to the $<111>$- axis of tungsten crystals, measurements of some shower
characteristics were already aimed to the development of a crystal-assisted positron
source. Theoretical estimations performed in \cite{BS02} display a rather good agreement
with results of recent experiments [10-13]. So, we can rely on our understanding of the
physics of shower formation and on numerical results, at least for tungsten crystals in
the energy range of incident electrons below 10 GeV. Note that just this energy range is
proposed in future linear collider projects ( 2, 6.2, and 10 GeV correspondingly for CLIC
\cite{CLIC},NLC \cite{NLC}, and JLC \cite{JLC} ) and is considered here.

Let us define the positron yield as the number of accepted positrons per one incident
electron and the optimal target thickness as that providing the maximal yield. According
to \cite{BKS95}, \cite{Artru94}, \cite{BS99}, the maximal yield from a crystal target is
always higher than that from an amorphous one and the excess goes up when the electron
energy increases. However, the magnitude of such an enhancement is small, less than
14$\%$ even at 10 GeV. The more pronounced advantage of crystal targets appear in a
considerable (by a factor of two at 10 GeV) decrease of the energy deposition. Indeed,
the thermal effects caused by the energy deposited in a target are a critical issue for
any powerful positron source based on the conventional scheme. We dwell mainly on this
issue in the present paper. Below qualitative arguments are given explaining the lower
energy deposition in crystals. The total deposited energy and the distribution of its
density over the target volume are calculated for crystal and amorphous tungsten targets
using the parameters of CLIC, NLC , and JLC . Thereby, a possible gain for these projects
resulting from the use of crystal targets in the positron source is estimated. For
accurate studies of thermal effects, some improvements have been performed in the
computer code developed in \cite{BKS95}, \cite{BS99}. The updated version of the code is
used to study both crystal and amorphous cases.

\section{Energy deposition in crystal and amorphous targets}

In the energy range under consideration we are dealing with a "soft"(see \cite{BKS87})
shower when pair production is entirely due to the conventional Bethe-Heitler mechanism,
while the crystal structure reveals in a considerable enhancement of the radiation
intensity and a softness of the photon spectra. Remember that this enhancement decreases
when the particle energy does so as the shower develops. Starting with some depth $L_0$
(see discussion in \cite{BKS95}, \cite{BS99}), further development of the shower proceeds
more or less in the same way for any (crystal or amorphous) type of the remaining part of
the target. For the sake of simplicity, calculations are performed here for the
all-crystal targets. However, they may serve as a good estimate for hybrid targets of the
same total thickness and with a crystal-part length of the order of $L_0$. Let us remind
that a hybrid target consists of a photon radiator made of a crystal followed by a pair
converter made of an amorphous piece.
\begin{figure}[h]
\centering
\includegraphics[width=0.48\textwidth
]{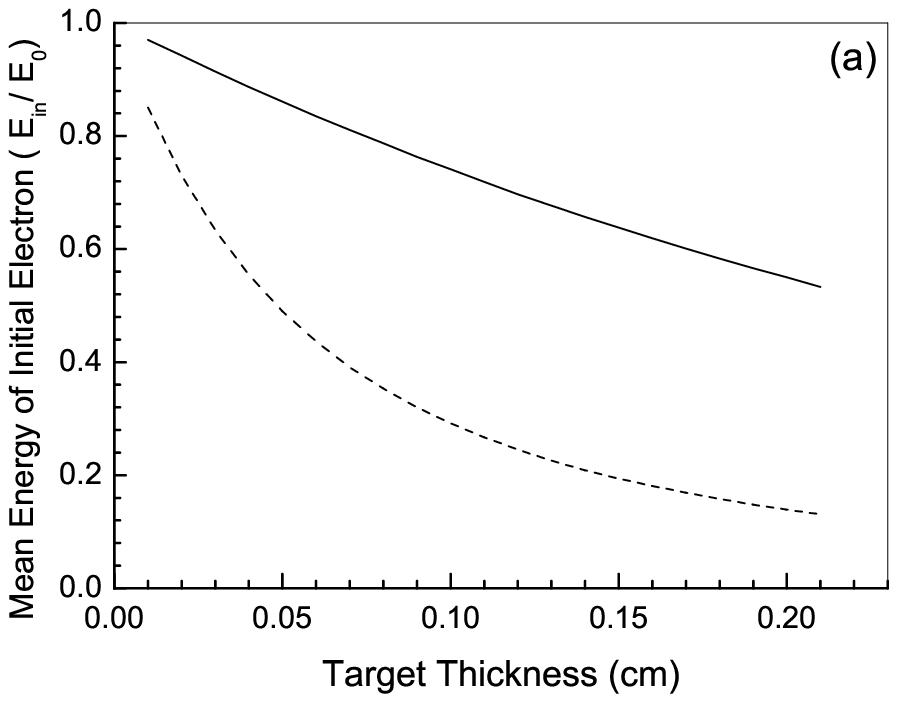}\hspace{0.025\textwidth}
\includegraphics[width=0.48\textwidth
]{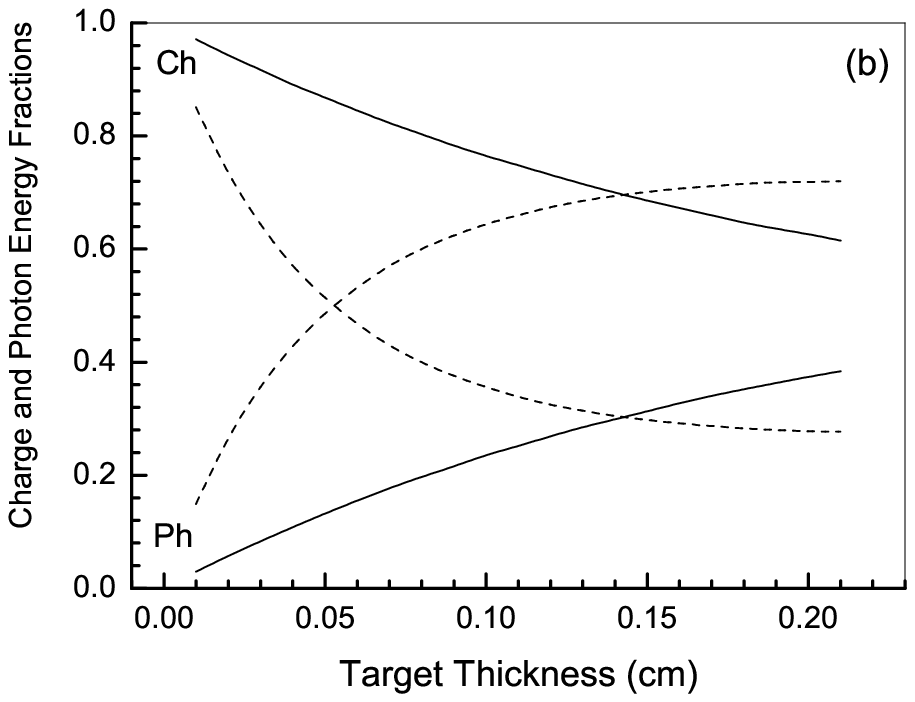} \caption{Mean energy of the initial electron in units of the incident beam
energy, $E_0 = 10~GeV$ (a) and : fractions of the total energy carried correspondingly by
all charged particles and photons (b). Solid lines are for amorphous and dashed for
crystal tungsten target. The incident beam is directed along the $<111>$-~axis of the
crystal. } \label{Fig:enbal1}
\end{figure}
From Fig.\ref {Fig:enbal1}, a value of $L_0 \gtrsim 0.2 cm (0.57 X_0)$ can be chosen for
10-GeV electrons, since the fraction of the total energy carried by photons ($\sim 0.72
$) has been already saturated at this depth and the mean energy of the primary electron
is sufficiently low to eliminate crystal effects in the last part. Such a saturation
takes place in amorphous targets as well, but with a lower conversion level ($\sim 0.59
$) and at substantially larger depth ($\sim 5L_0$ at 10 GeV). Only a small part (less
than 0.4$\%$ in the above example) of the beam power is deposited over $L_0$ and the
energy-deposition density is much less (about 8 times at 10 GeV) than its peak value. So,
the crystal part of a hybrid target is not exposed to serious thermal effects which
appear at larger depths in the later stage of the shower development.

\begin{figure}[h]
\centering
\includegraphics[width=0.48\textwidth
]{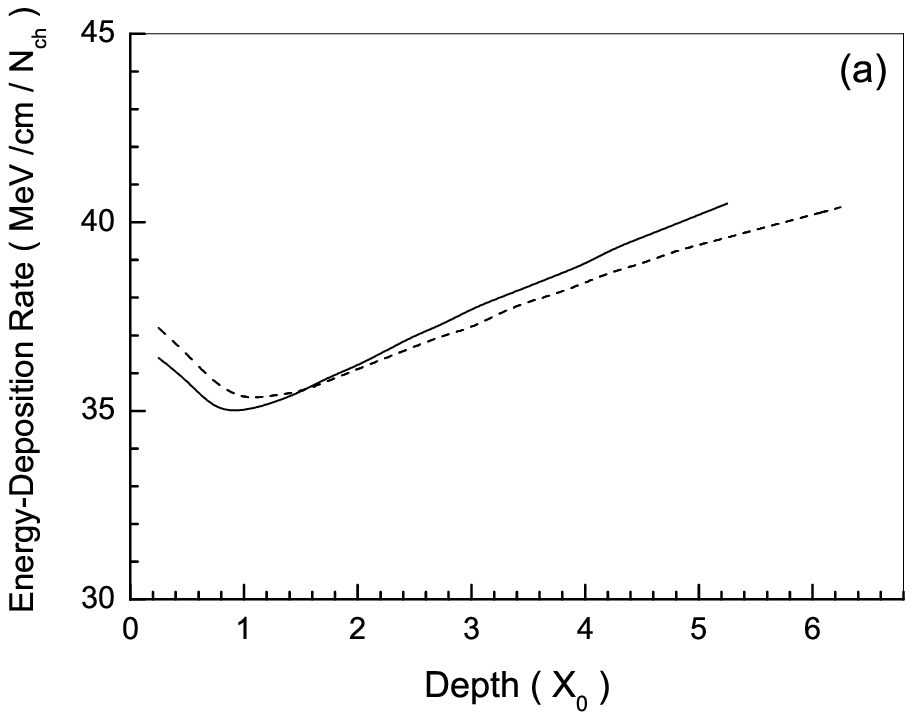}\hspace{0.025\textwidth}
\includegraphics[width=0.48\textwidth
]{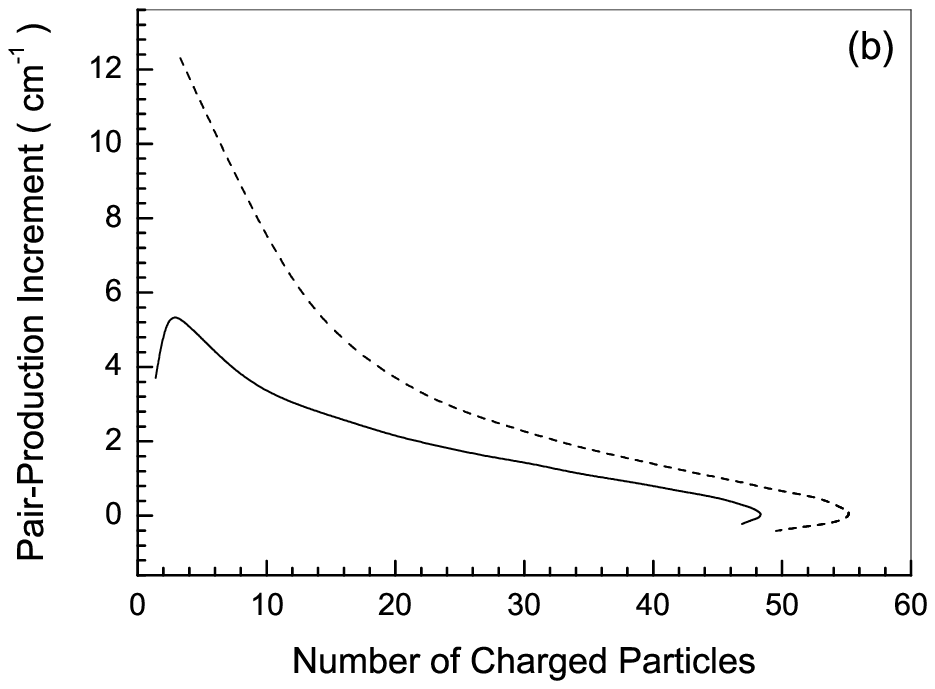} \caption{Energy deposition rate per charged particle (a) and the
logarithmic increment  (b) in tungsten targets at $E_0 $ =10~GeV . Solid lines are for
amorphous and dashed for crystal targets.} \label{Fig:rate2}
\end{figure}

From calculations performed in \cite{BKS95}, \cite{BS99}, the energy deposition in
equivalent (providing the same positron yield) targets is always less in the crystal
case. Let us present some qualitative arguments to explain this fact. The main process
leading to the energy deposition is the ionization loss, the rate of which, $q(z)$, reads
approximately as $q(z)\simeq C_Q\cdot N_{ch}(z)$, where $N_{ch}(z)$ is the number of
charged particles at the depth $z$. Strictly speaking, the coefficient $C_Q $ may depend
on $z$ but its small variation as well as a small difference of $C_Q $-values in crystal
and amorphous cases are neglected in our estimation. So, the total energy, $Q(L)$,
deposited over the thickness $L$ reads
\begin{equation}\label{qen1}
Q(L)\,=\,\int\limits_{0} ^{L}dz \, q(z)\,\simeq C_Q \int\limits_{0}^{L}dz \ N_{ch}(z)
\,\,,
\end{equation}
or, going over to the variable $N_{ch}(z)$
\begin{equation}\label{qen2}
Q(L)\,\simeq C_Q \int\limits_{1}^{N_{ch}(L)}dN_{ch}\Bigl(\frac{d\ln
N_{ch}}{dz}\Bigr)^{-1} \,\,.
\end{equation}
For sufficiently large $L$, the positron yield is roughly proportional to the total
number of charged particles, $N_{ch}(L)$, i.e.,for equivalent targets, the integrals in
(\ref{qen2}) are taken over the same region of variable $N_{ch}$ in both cases. To prove
our statement, it remains only to verify that the logarithmic derivative which appears in
the denominator of the integrand in (\ref{qen2}) is larger for crystals. This derivative,
or logarithmic increment, characterizes the growth of the number of charged particles. As
seen in Fig.\ref {Fig:enbal1}, the conversion of the initial electron energy into photons
is going faster for crystals, where, correspondingly, the pair-production process starts
earlier and is more intensive, resulting in a larger increment. For the purposes of
illustration, the energy-deposition rate per charged particle and the logarithmic
increment are shown in Fig.\ref {Fig:rate2}. We emphasize that the energy-deposition rate
is
practically the same for crystal and amorphous samples, being almost independent of the
electron energy. The non-constancy of this rate is mainly due to the contributions of
other processes like photon absorbtion and annihilation of positrons which are taken into
account by our simulations but were ignored in the estimation of $q(z)$ used in
(\ref{qen1}). Evidently, the role of these processes increases with growing depth. A slow
decrease of the rate at comparatively small depth is due to that of the mean particle
energy. The point is that we use the so-called non-restricted energy loss description,
where the rate diminishes when the particle energy does so. On the whole, the results of
simulations presented in Fig.\ref {Fig:rate2} confirm the above qualitative
considerations.

\begin{figure}[h]
\centering
\includegraphics[width=0.48\textwidth
]{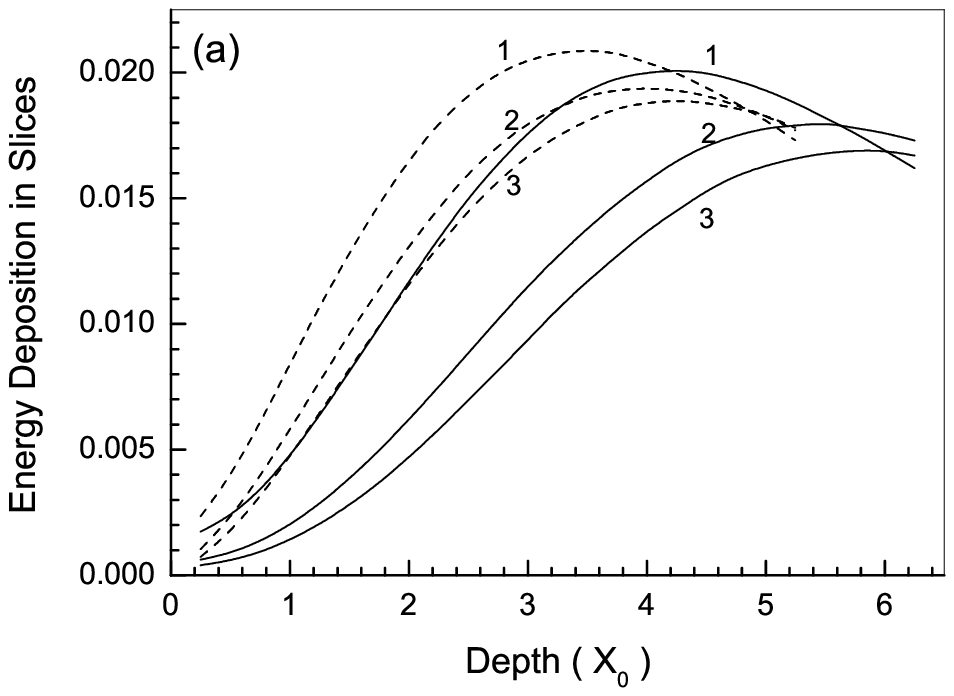}\hspace{0.025\textwidth}
\includegraphics[width=0.48\textwidth
]{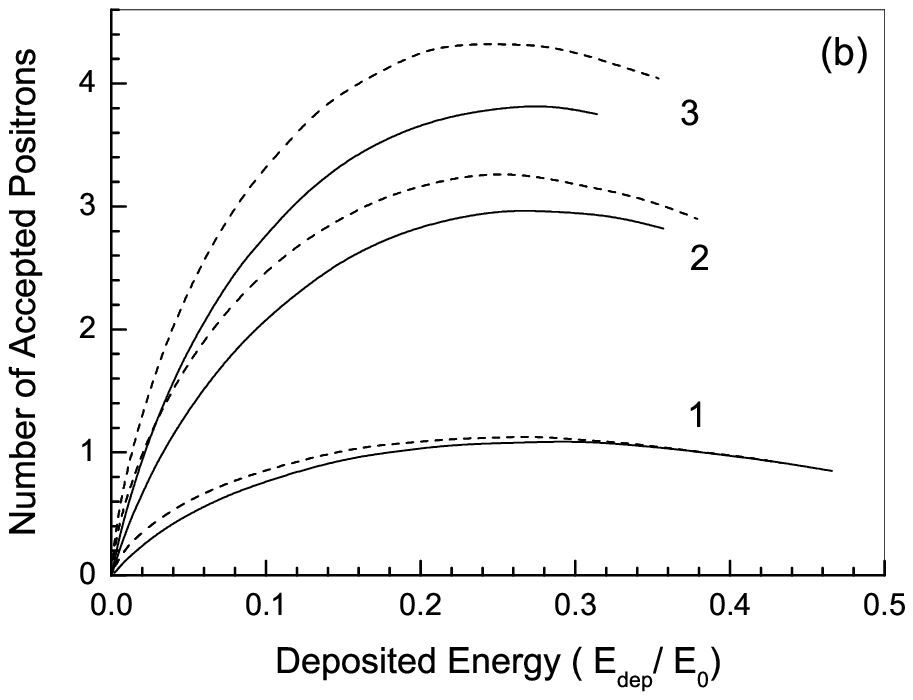} \caption{Fraction of energy $\Delta E_{dep} (z)/E_0$ deposited in $z$-slices
(a) and number of accepted positrons as a function of the deposited energy $E_{dep} /E_0$
(b) at $E_0$ = 2~GeV ( curves 1 ), 6.2~GeV ( curves 2 ),and 10~GeV ( curves 3 ). Solid
lines are for amorphous and dashed for crystal targets. } \label{Fig:edep3}
\end{figure}

As shower develops, particles of sufficiently low energy may substantially change their
direction of propagation due to single or multiple scattering. So that there even appears
a "backward" flux of particles moving upstream; they do not increase the positron yield,
however, they heat the target. This contribution is taken into account in our
simulations, where we are able to trace separately effects from "forward" and "backward"
particles and photons. To study the shower characteristics depending on depth ($z$), the
target is divided into slices by planes perpendicular to the incident beam direction,
taken as $z$-axis. The spacing $dz=0.25\,X_0$ is used (remember that $X_0$ =3.5mm for
tungsten). Energy deposition is simulated within each slice and various distributions are
recorded at the right-hand boundary ( plane ) of the slice. In particular, the
development of momentum distributions for positrons and photons is obtained as well as
that of the beam spot size. Shown in Fig.\ref {Fig:edep3}(a) is the energy, $\Delta
E_{dep} (z)$, deposited in slices in units of $E_0$. This quantity is higher for the
lower energy. Note that the
deposited power,$P_{dep}$, reads as $P_{dep}=P_0\cdot E_{dep}/E_0$ where $P_0$ is the
incident beam power. The contribution of the "backward" particles to $E_{dep}$ increases
with the target thickness $L$; it amounts typically to about 20$\%$ at $L\sim 4 X_0$ and
thereby is not negligible. The results of simulations presented in Fig.\ref {Fig:edep3}
(b) clearly confirm once more (cf. Fig.9 in \cite{BKS95}) the statement concerning the
comparison between the values of $E_{dep}$ in the equivalent targets. The positron yield
is calculated using the "theoretical" acceptance conditions from \cite{CLIC} ( see
Figs.11 and 12b in \cite{CLIC} ) in all three cases. The angular spread of the incident
beam is neglected and the transverse size of the beam, $\sigma (\sigma_x=\sigma_y= \sigma
)$ is set to 1.6~mm at 2 and 6.2 GeV, and to 2.5~mm at 10 GeV. At $E_0$=2 GeV, a
simulation for $\sigma$=2.0~mm was performed as well. Corresponding results, not shown on
graphs, are presented in the Table below.

\begin{figure}[h]
\centering
\includegraphics[width=0.48\textwidth
]{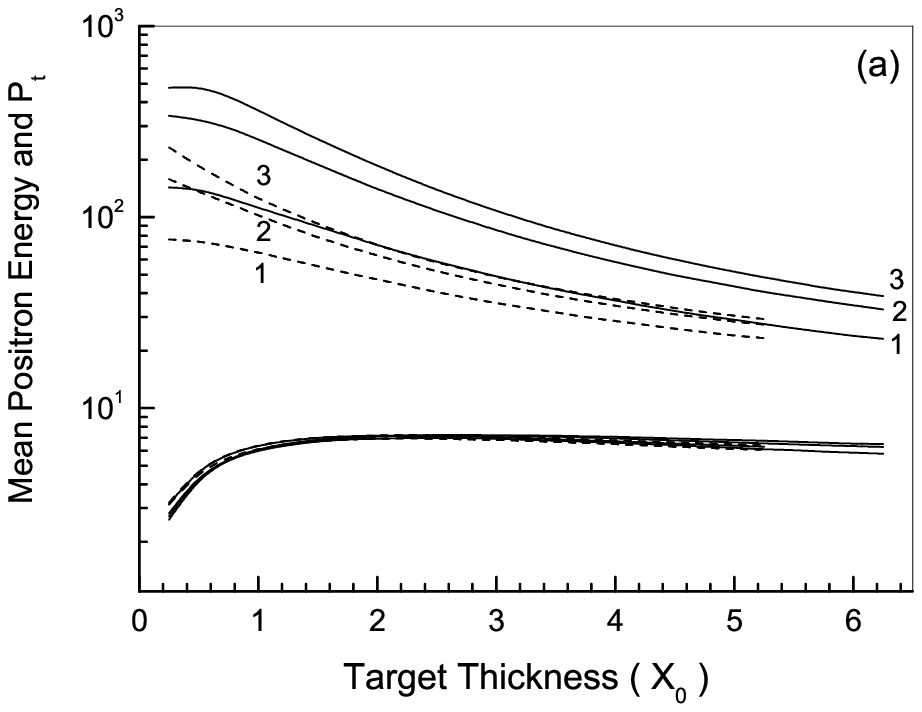}\hspace{0.025\textwidth}
\includegraphics[width=0.48\textwidth
]{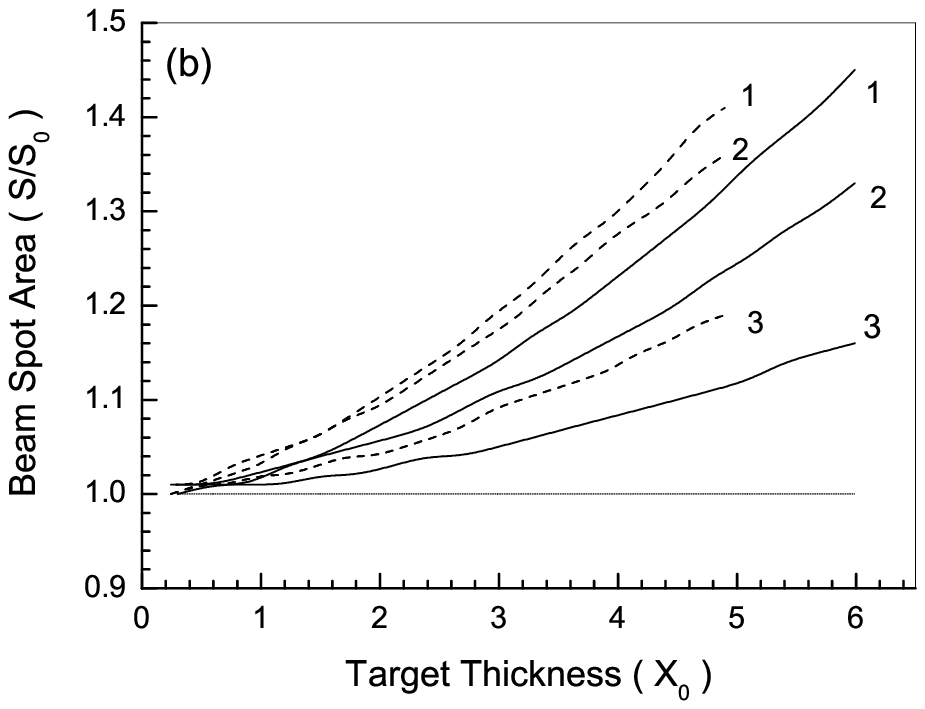} \caption{(a) : Mean positron energy  (in MeV, upper curves) and transverse
momentum  (in MeV/c, lower curves); (b) : beam spot area development. Incoming energies
are $E_0$ = 2~GeV ( curves 1 ), 6.2~GeV ( curves 2 ),and 10~GeV ( curves 3 ). $S_0 $
corresponds to the incident beam. The straight line $y=1$ on (b) is drawn to guide the
eye. Solid lines are for  amorphous and dashed for crystal targets.} \label{Fig:menspo4}
\end{figure}

Let us remind now that, at equal depths and initial energies, charged particles are
softer and have a larger angular spread for crystal targets (see, e.g., Figs. 2-5 in
\cite{BS02} and corresponding discussion). All other things being equal, positron spectra
are softer at a lower initial energy. As an illustration of these features, the mean
energy and the transverse momentum, $<p_t>$, of "forward" positrons are shown in Fig.\ref
{Fig:menspo4}(a). We emphasize that, starting with $L \sim X_0$, $<p_t>$ is almost
constant and practically independent of the initial electron energy and the type of the
target (note the merging of six different curves on Fig.\ref {Fig:menspo4}(a)).
Such result confirms and extends those concerning an amorphous target; it is essentially
due to a counterbalance between the increase of the angle as a consequence of the
multiple scattering and the decrease of the positron energy with increased thickness
\cite{Artru94}. At the same time, the larger angular spread leads in the crystal case to
a larger beam spot area, $S$, as seen in Fig.\ref {Fig:menspo4}(b) where $S/S_0$ is
plotted for "forward" charged particles. For evident reasons, the spot area of the
"backward" charged flux ( not shown in Fig.\ref {Fig:menspo4}) is somewhat larger
(typically by 20-25 $\%$) than that of the "forward" one.

Whereas the average deposited power can be handled somehow (e.g., by rotating the target
and removing the excess heat through water cooling as used at SLC), the local and nearly
instantaneous energy deposition is unavoidable, while being of critical concern for
target damage. Basing on the analysis of the SLC damaged target ( see \cite{LANL}), it is
now adopted that the peak energy-deposition density should not exceed 35 J/g to ensure a
sufficiently long term of safe operation. In our simulations, the total scanned volume is
a cylinder coaxial with the incident beam direction. The radius, $R$, of the cylinder was
about $R\simeq 3.7\sigma$, in which case less than one thousandth of initial electrons
does not hit the cylinder at entry. This cylinder is sliced into disks of the thickness
$dz=0.25\,X_0$, i.e., $z$-spacing is the same as in above calculations, allowing a mutual
checking. In turn, each disk is divided by circles of uniformly increasing radii with the
step $dr=0.02 R$ into 50 parts - one internal disk (altogether, such disks form the
internal cylinder ) and 49 rings. More precisely, we have $dr(\sigma$=1.6mm)=0.12mm,
$dr(\sigma$=2.0mm)= 0.15 mm, and $dr(\sigma$=2.5mm)=0.185mm. The energy deposition and
the number of charged particles is simulated in each meshed volume providing
corresponding transverse distributions for each $z$-slice. It is noteworthy that
$z$-dependencies of shower characteristics derived from 3-dimensional distributions
coincide with those obtained in direct calculations, thereby verifying the consistency of
the two essentially different forms of the output which we have used.
\begin{figure}[h]
\centering
\includegraphics[width=0.48\textwidth
]{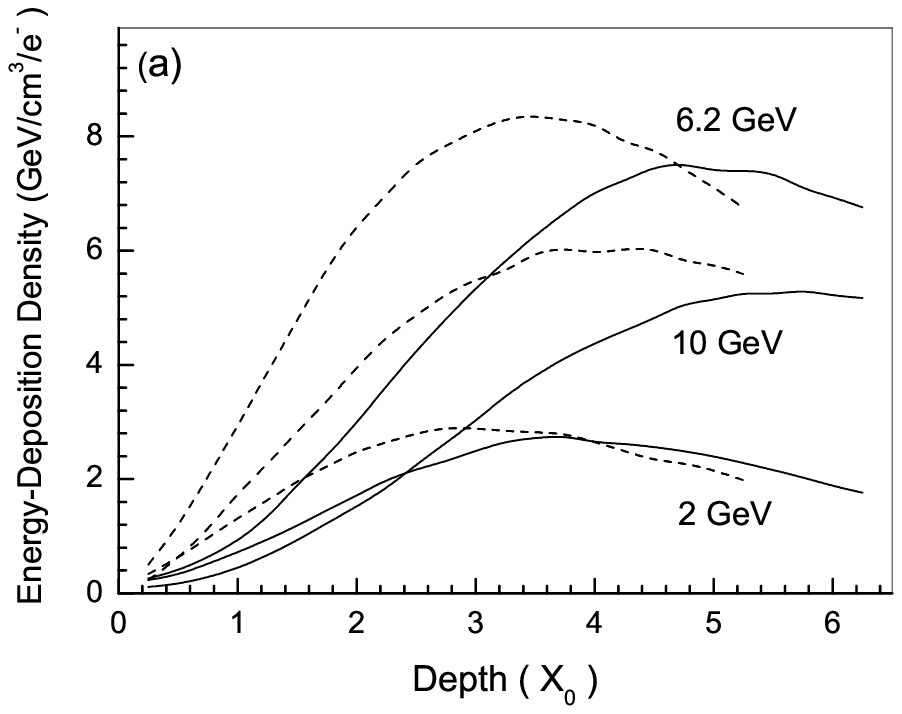}\hspace{0.025\textwidth}
\includegraphics[width=0.48\textwidth
]{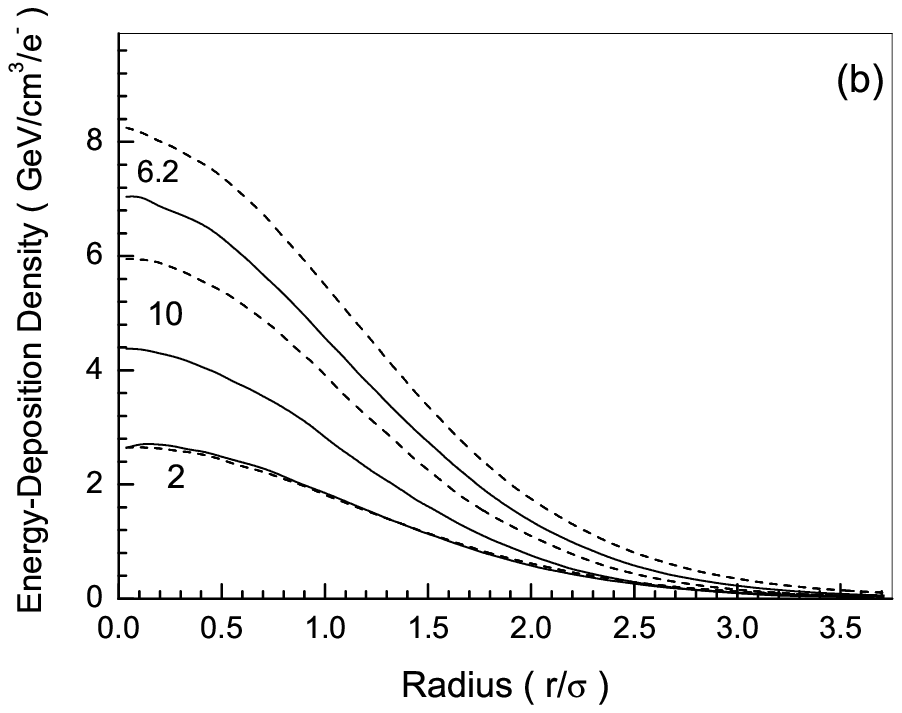} \caption{(a): Energy-deposition density in the internal cylinder, versus
the  depth, at indicated initial energies; (b): Transverse distribution of the
energy-deposition density at $z =4.0 X_0$. Solid lines are for  amorphous and dashed for
crystal targets. } \label{Fig:endede}
\end{figure}
From Fig.\ref{Fig:endede} (b), the energy-deposition density (EDD) drops at large cell
radius, being maximal in the internal disk of each slice. Shown in Fig.\ref{Fig:endede}
(a) is the EDD per one incident electron in the internal cylinder, versus the depth. Like
the positron yield and the total number of charged particles , the peak values of EDD
(PEDD) in crystals are somewhat higher and are reached at smaller depths, compared with
amorphous targets. At equal beam size, PEDD is higher for the higher energy (cf. pairs of
curves in Fig.\ref{Fig:endede}(a) calculated at 2 and 6.2 GeV for the same $\sigma=$1.6
mm). However, by increasing the beam size, lower values of PEDD may be obtained even at
higher energy. This is seen in Fig.\ref{Fig:endede} (a) if we compare the curves
calculated at 10 GeV ( $\sigma$ =2.5 mm) with those at 6.2 GeV. Examples of transverse
distributions of EDD are plotted in Fig.\ref{Fig:endede}(b) for the same depth $z=4 X_0$.
The distribution of charged particle density (not shown) is similar to that of EDD with a
ratio EDD/density being almost constant within the first ten cells and roughly equal to
the energy-deposition rate. In Fig.\ref{Fig:endede}(b) this similarity reveals in larger
width of distributions for crystal targets (cf. Fig.\ref{Fig:menspo4}(b) ). The EDD is
presented in Fig.\ref{Fig:endede} in units of $GeV\cdot cm^{-3}$ which for tungsten
corresponds to $1\,GeV\cdot cm^{-3}=8.30\cdot 10^{-12} J/g $.

Due to an extremely short duration of the pulse, contributions to EDD from all the
incident electrons are added and the resulting EDD-value is simply the product of the
obtained EDD-value per $e^-$ and the number of electrons per pulse, provided that the
target reverts to initial thermal conditions
\begin{table}[h]
\begin{center}
\caption{{\bf Energy deposition in crystal and amorphous targets}. The fraction of an
incident beam power deposited in a target, $R_{th}=P_{dep}/P_0$; the peak energy
deposition density ,$ PEDD $, in units of J/g; the energy,$E_0$, and transverse size of
an incident beam,$ \sigma $; target thickness, $L$, measured in conventional radiation
lengths, $X_0$; gain in the total deposited power, $ G=(1-E_{cr}^{dep}/E_{am}^{dep})\cdot
100$, from comparison of $ W_{cr} $ and $ W_{am} $ targets } \label{Tab:compar} \vskip
.5cm
\begin{tabular}{|c|c|c|c|c|c|c|c|c|c|} \hline
Beam & \multicolumn{3}{|c|}{$E_0=$2 GeV,$\sigma=$2.0 mm}&
\multicolumn{3}{|c|}{$E_0=$6.2
GeV,$\sigma=$1.6 mm} & \multicolumn{3}{|c|}{$E_0=$10 GeV,$\sigma=$2.5
mm} \\
\hline Target & $W_{75}Re_{25}$ & $W_{am}$ & $W_{cr}$ & $W_{75}Re_{25}$
& $W_{am}$ &
$W_{cr}$ & $W_{75}Re_{25}$ & $W_{am}$ & $W_{cr}$
\\ \hline $L$ & 4.0 & 4.0 & 3.0& 4.0 & 4.0 & 2.5& 6.0 & 6.0 & 3.0
\\ \hline $R_{th}$ & 0.248 & 0.238 & 0.193 & 0.142 & 0.147 &
0.106&0.310&0.291&0.137
\\ \hline $PEDD$ &35.0&32.3&33.8&35.0&56.1&60.3&35.0&28.2&30.6 \\ \hline
$G$& \multicolumn{3}{|c|}{19$\%$}&
\multicolumn{3}{|c|}{28$\%$} & \multicolumn{3}{|c|}{53$\%$} \\
\hline
\end{tabular}
\end{center}
\end{table}
during the repetition period. Some results concerning the energy deposition in crystal
and amorphous targets are presented in Table \ref{Tab:compar}. For the sake of
comparison, the thicknesses of amorphous tungsten targets ($W_{am}$) are the same as
proposed for $W_{75}Re_{25}$ targets in the projects \cite{CLIC}, \cite{NLC}, and
\cite{JLC}. For crystal targets ($W_{cr}$), the corresponding thicknesses are determined
from the equivalence rule (the same positron yield as for $W_{am}$ target). Our values
for $PEDD$ given in Table \ref{Tab:compar} are obtained using the same numbers of
electrons per pulse, $N_e(2\,GeV)=2.08 \cdot 10^{12} $, $N_e(6.2\,GeV)=0.96 \cdot 10^{12}
$, and $N_e(10\,GeV)=0.64 \cdot 10^{12} $, as in \cite{CLIC}, \cite{NLC}, and \cite{JLC}
respectively.

Since the parameters of the two amorphous targets such as radiation lengths and densities
almost coincide, the same is expected for shower characteristics at equal depths.
However, one should bear in mind that two different computer codes were used in
simulations, EGS4 for $W_{75}Re_{25}$ targets, and our code for $W_{am}$ targets. From
Table \ref{Tab:compar}, the concordance of the two approaches is almost perfect for the
fraction of the deposited power, $R_{th}$. Concerning the peak energy-deposition density,
$PEDD$, the values obtained for the $W_{am}$ target at 2 and 10 GeV are somewhat smaller
than those for the $W_{75}Re_{25}$ target and the distinctions are not too big. On the
contrary, our value for the $W_{am}$ target is substantially larger at 6.2 GeV. Let us
argue the point in detail.  At equal beam size, the peak EDD/electron is expected to be
roughly proportional to the initial electron energy, $E_0$. This assertion follows from
the qualitative consideration performed above and is verified by calculations. For
example, if we compare  the curves in Fig.\ref{Fig:endede}(a) calculated at 2 and  6.2
GeV for the same $\sigma=$1.6 mm, we obtain $K=$2.7 as the peak EDD ratio instead of
$K=$3.1 from the rough estimate where $K(E_{02},E_{01})=E_{02}/E_{01}$. At $L = 4X_0$,
which corresponds to NLC conditions, this ratio further diminishes up to 2.55 since the
peak EDD value is achieved for  $E_0$ =6.2 GeV at a larger depth. So, at equal beam size,
the relationship between $PEDD$ values at different energies reads roughly as
\begin{equation}\label{est3}
PEDD(2)\simeq PEDD(1)\cdot K(E_{02},E_{01})\cdot N_e(E_{02}) / N_e(E_{01})\,\,.
\end{equation}
The $PEDD$ in $W_{75}Re_{25}$ target at $E_0$ =2 GeV  was calculated in \cite{CLIC} not
only for $\sigma=$2.0 mm, but also for $\sigma=$1.6 mm, where the value of 53.1 J/g has
been obtained ( at these conditions we have 47.7 J/g for the $W_{am}$ target). Using the
estimate (\ref{est3}) with $PEDD(1)$=53.1 J/g and $K=$2.55, we obtain for NLC conditions
$PEDD(2)$=62.5 J/g which is consistent with our result for the $W_{am}$ target.
Conversely, starting with results of \cite{JLC} for the $W_{75}Re_{25}$ target, we obtain
the estimate ($PEDD(1)$=35 J/g, $E_{01}$= 10 GeV, $K=$0.62) $PEDD(2)\simeq$ 33 J/g at
$\sigma=$2.5 mm, $E_{02}$=6.2 GeV. So, the $PEDD$ value of 35 J/g at $E_0$=6.2 GeV is
more consistent with the beam size of $\sigma=$2.5 mm than with $\sigma=$1.6 mm indicated
in \cite{NLC}.

Comparing the magnitude of thermal effects in $W_{am}$ and $W_{cr}$ targets providing the
same positron yield, we conclude that , using crystal targets, the total deposited energy
can be considerably diminished while the peak value of the energy-deposition density is
kept approximately on the same level as in the amorphous case.

\vspace{0.25 cm} \noindent{\bf Acknowledgements} \vspace{0.25 cm}

\noindent One of us (V.S.) is thankful for kind hospitality during his stay at LAL where
a part of this work has been done. He is also grateful to the Russian Fund of Basic
Research for partial support of this work by Grants 01-02-16926 and 03-02-16510.


\end{document}